\documentclass[a4paper,11pt]{article}
\usepackage[dvips]{graphicx}
\usepackage[lflt]{floatflt}
\def \beq{\begin{equation}}
\def \eeq{\end{equation}}
\def \I{{\it i}}

\begin{document}
\author{P. Buonsante\footnote{{\tt buonsant@fis.unipr.it}}, R. Burioni\footnote{{\tt burioni@fis.unipr.it}} and  D. Cassi\footnote{{\tt cassi@fis.unipr.it}}}
\title{Topological Reduction of Tight-Binding Models on Complex Networks}
\date{}
\maketitle
\begin{center}
{\it Dipartimento di Fisica, Universit\`a degli Studi di Parma and Istituto Nazionale per la Fisica della Materia, Unit\`a di Ricerca di Parma, Parco Area delle Scienze 7/a - 43100 Parma, Italy}
\end{center}
\begin{abstract}
 Complex molecules and mesoscopic structures are naturally described by general networks of elementary building blocks and tight-binding is one of the simplest quantum model suitable for studying the physical properties arising from the network topology. Despite the simplicity of the model, topological complexity can make the evaluation of the spectrum of the tight-binding Hamiltonian a rather hard task, since the lack of translation invariance rules out such a powerful tool  as Fourier transform. In this paper we introduce a rigorous analytical technique, based on topological methods, for the exact solution of this problem on branched structures.
Besides its analytic power, this technique is also a promising engineering tool, helpful in the design of netwoks displaying the  desired spectral features.
\end{abstract}
\pagebreak
\section{Introduction: The Role of Topology}
Physical systems like complex molecules and mesoscopic structures  consist of a large collection of elementary building blocks, atoms or  
small molecules, connected to give a more  or less  ordered structure.
When a regular and periodic pattern  is present, as in crystals, it is
possible to isolate an elementary  cell from which all the features of
the  system  stem.  At  length scales  much higher than  the  cell's
size  these features  depend uniquely  on the  dimensionality
 of the cell's  packing. Periodicity is particularly useful in
the  analysis of  the  spectral  properties of  the  system: it  makes
it possible to introduce  a (pseudo) wave vector,  $\bf{k}$, which
labels every quantity of interest,  such as the energy or the Green's
functions \cite{latticesB}. 
When, on the other  hand, no periodic
pattern is recognizable, such a simple picture does not apply.  In
 this paper we will  develop an analytical
method  fit for the  study of the  spectral properties of  a wide class of 
discrete structures.  We  will  focus  on the  simplest quantum
 model, namely  the
tight-binding  model,  defined  on  structures lacking  the  regular
arrangement  typical of  perfect crystals.   In order  to  analyze the features  strictly arising from the topology of the
arrangement  we  will  simplify  as  much as  possible  the  "internal
structure" of the building blocks.  The resulting model is simple yet
rich, and  it displays all  the qualitative effects due to topology.
In this framework we will illustrate an analytical technique, which we
call  {\it  bud-reduction},  useful  for a  large  class  of  branched 
networks.  As  its name  suggests it makes it possible to account the
spectral  features stemming  from an  entire substructure  of  a given
discrete network  by reducing it to  what we call its  {\it bud}. This
operation is a sort of  topological renormalization, in the sense that
the presence of an  entire substructure is accounted by means
of an  on-site potential  of geometrical origin.   Moreover it  can be
performed recursively, this meaning that a structure can be reduced to
its bud  either as  a whole,  or after some  of its  substructures have
already undergone bud-reduction.  From  a {\it top-down} point of view
this  technique allows  a geometrical  simplification of  the original
network  and a  direct understanding  of the  effect  of the  network
topology  on its  spectral properties.   On  the other  hand it  seems
particularly  well  suited for  the  engineering of  spectral
properties: once the desired spectral structure has been identified it
can  be obtained  by  means  of an  appropriate  network design.   

The present paper  is organized as follows:  Section \ref{graphsS} briefly
deals with the mathematical  representation of discrete networks, from
the topological point of view.  A number of useful definitions are also
given  there.   Section  \ref{tbS}  contains an  introduction  to  the
tight-binding model, generalized to  the case of an arbitrary 
network.  In Section \ref{ldosS}  we introduce some mathematical tools
which allow  the computation of  the spectrum of such an arbitrary network,
independent of its symmetries and  regularities.  In  Section
\ref{brS} we describe  bud-reduction and provide  numerous
 examples which clarify the technique and its application. Section 
\ref{conclS} consists in our conclusions. Section \ref{appx} is devoted to some detailed calculations.
\section{Representation of an Arbitrary Discrete Network : Graphs}
\label{graphsS}
When  no periodicity  is present  lattice cannot be used  to describe discrete structure,  the appropriate choice being graphs
\cite{graphsB}.  A  graph,  ${\cal  G}=\{\Sigma, \Lambda  \}$,  consists
 of a  set of points, or sites, $\Sigma$,
and  a set  of links,  $\Lambda$,  connecting  points pairwise. In  the
present paper lowercase italic  letters $i$, $j$, $k$,..., will denote
the sites  of a  network, and  pairs of  lowercase letters
enclosed  in parentheses, $(h,k)$,  will denote  the link  joining the
relevant  sites.  From  an algebraic  point of  view  a  graph  is
completely   described  by   what   is  called   its  {\it   adjacency
matrix}. Every entry  of this matrix corresponds to  a couple of sites
of the network, and  it equals one if and only if this couple  is joined by a
link, otherwise it is zero: 
\beq
\label{adj_m}
A_{i\,j}=\left\{
\begin{array}{ll}
1 & {\rm if\;}  (i,\,j) {\rm \;is\;a\;link\;of\;the\;graph}\\ 0 & {\rm
otherwise}
\end{array}\right.
\eeq       
Any        sequence       of       consecutive       links,
$(i,\,h),(h,\,k),...,(l,\,m),(m,\,j)$, is referred  to as a {\it path}
on the graph. In the following  a path starting at site $i$ and ending
at site  $j$ will be denoted  by the symbol ${\cal P}_{i\,j}$. 
 In general  such  a path  may reach  its
ending site, $j$, also at an intermediate step. When the path never passes through its ending site at an intermediate step it will be denoted by the symbol
${\cal F}_{i\,j}$. 
\begin{figure}[h]
\begin{center}
\includegraphics[width=16cm]{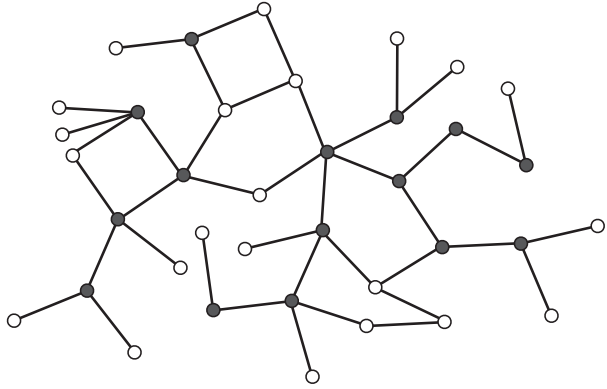} 
\caption{Graph of a generic network. Filled circles denote branching sites, i.e. sites which are the unique intersection of any two connected subgraphs. Unfilled circles denote non-branching sites.}
\label{branchF}
\end{center}
\end{figure}
The  bud-reduction
technique we will illustrate in  Section \ref{brS} can be performed at
sites of the graph we will call  {\it branching sites}. A site $*$ of a
graph ${\cal G}$ is a branching site if it is the only intersection of
any  two  connected subgraphs ${\cal  B}_1$,  ${\cal  B}_2$  of ${\cal  G}$;  in mathematical terms  
\beq
\label{branches}
{\cal G}=\{\Sigma,\,   \Lambda   \},\quad   {\cal B}_i=
\{\Sigma_i,\,  \Lambda_i \},\quad \Sigma=\Sigma_1  \cup \Sigma_2,
\quad   \Lambda=\Lambda_1 \cup \Lambda_2,\quad \left\{*\right\}
=\Sigma_1  \cap   \Sigma_2,  \quad  \emptyset=\Lambda_1  \cap \Lambda_2  
\eeq 
Any  two  subgraphs  joined by  a  branching site  are
referred to as {\it branches}.   What makes a branching site peculiar is
the fact that  any path joining sites on  different branches is forced
to pass through it. Figure~\ref{branchF}.  illustrates the concept of branching site.
\pagebreak
\section{Tight-Binding on a Generic Structure}
\label{tbS}
The tight-binding approximation \cite{latticesB} is widely used when dealing with 
quantum particles on discrete structures, such as solids, molecules,
atomic clusters. The Hamiltonian of such a system has the form 
\beq
H =\frac{{\bf p}^2}{2\,m}+\sum_i V_i({\bf r})
\eeq
where $V_i(\bf r)$ is the potential energy due to the interaction with the
$i$-th atom of the structure. Assuming that the wavefunctions of the Hamiltonian
may be expressed as linear combinations of atomic orbitals, that each single  atom of the structure contributes with a single $s$-type orbital,  and that these orbitals 
in turn have a non-zero overlap only if they come from adjacent atoms,  the resulting
 Hamiltonian has the matrix form \cite{gaussB}
\beq
\label{GTBH}
H^{\rm TB}_{i\,j} = t \, A_{i \, j} + a z_i \delta_{i \, j}
\eeq
It consists of an off-diagonal  term, called {\sl hopping term}, and of
a diagonal  term, called {\sl local  term}. The hopping term contains  
the adjacency matrix describing the topological arrangement of the 
structure and the so-called {\it hopping parameter}, $t$.
The local term contains the {\it local parameter} and the coordination number, 
$z_i=\sum_j A_{i\,j}$, namely the number of nearest neighbours of the $i$-th
site.  
On a  periodic lattice (with homogeneous elementary cell) this term  is site-independent and  it can  be dropped  without  loss of
generality,  since  it simply results in  a  rigid shift  of  the
spectrum.   When  the geometric  structure  is  non-periodic this  cannot be done, and  moreover there is  no natural tool
such as the Fourier trasform for the diagonalization of the Hamiltonian.
Hamiltonians of the form (\ref{GTBH}) are widely used. They also apply to more general systems than the one described above, in which for instance the building blocks at each site are not simple atoms but more complex structures\cite{dend1,dend2}.  
In spite of the approximations which it has undergone, the model described above is nevertheless rich and capable of displaying the features arising from the topological arrangement of the structure.

\section{ Green's Functions and Local Density of States}
\label{ldosS}
A route to the spectrum of the Hamiltonian (\ref{GTBH}) alternative to direct diagonalization passes through the so-called {\sl Local Density of States} (LDOS), which is related to Green's functions by the formula \cite{eco}
\beq
\label{GF2LDOS}
\rho_i(\omega)=-\frac{1}{\pi} \lim_{\epsilon \to 0} {\rm Im} (G_{i \, i}(\omega + \I \epsilon))
\eeq
\noindent
Before going into further detail we point out that most of the quantities we will introduce carry an explicit site dependence, which in the case of perfect crystals with simple primitive cell usually drops, due to the spatial homogeneity.
The Green's function is defined by
\beq
\label{GF1}
G(\omega)= \left[\omega - H\right]^{-1} = \left[\omega - Z a - t A\right]^{-1}
\eeq
\noindent
where $Z_{i \, j} = z_i \delta_{i \, j} $. By using a complete set of eigenfunctions of the Hamiltonian operator 
\beq
H | k \rangle = \omega_k | k \rangle
\eeq
\noindent
one gets
\beq
\label{GF2}
G(\omega)= \sum_k \frac{|k\rangle \langle k|}{\omega - \omega_k} 
\eeq
\noindent
thus
\beq
\label{GF3}
G_{i \, i}(\omega)= \sum_k \frac{\| \langle i|k\rangle \|^2}{\omega - \omega_k}\eeq
\noindent
where $| i \rangle $ is the eigenfunction of the position operator at site $i$. Recalling that
\beq
\nonumber
\lim_{\epsilon \to 0}{\rm Im} \frac{1}{\omega \pm \I \epsilon} = \mp \pi
\eeq
\noindent
from (\ref{GF2LDOS}) one gets
\beq
\label{GF2LDOS2}
\rho_i(\omega)= \sum_k \delta(\omega-\omega_k) \| \langle i|k\rangle \|^2
\eeq
The last formula sheds light upon the significance of the LDOS at site $i$: it is a sort of projection of the density of states (DOS) on that site.
The DOS is recovered by simply summing the LDOS over all sites:
\beq
\label{LDOS2DOS}
\sum_i \rho_i(\omega)= \sum_k \delta(\omega-\omega_k) \sum_i \| \langle i|k\rangle \|^2 = \sum_k \delta(\omega-\omega_k) =  \rho(\omega)
\eeq
The normalization of the eigenfunctions of the position operator and the completeness of the Hamiltonian's eigenfunctions yield the normalization of LDOS:
\beq
\label{NormLDOS}
\int_{-\infty}^{\infty} \, d\, \omega \, \rho_i(\omega)= \sum_k \| \langle i|k\rangle \|^2 \int_{-\infty}^{\infty} \, d\, \omega \, \delta(\omega-\omega_k) =\sum_k \| \langle i|k\rangle \|^2 =  \langle i  |i\rangle  = 1
\eeq
\noindent
Simple algebraic manipulations of (\ref{GF1}) yield
\beq
\label{GF4}
G_{i \, j}(\omega)=  \frac{1}{\omega - z_j \, a}\sum_{k=0}^{\infty} \, \left[(\omega - Z \, a)^{-1}  A \,t \right]^k_{i \, j} = g_j \sum_{k=0}^{\infty} \, t^k \sum_{h_1, h2,..., h_{k-1} } \, g_i \, A_{i \, h_1} \, g_{h_1} \, A_{h_1 \, h_2} ... \, g_{h_{k-1}} \, A_{h_{k-1} \, j}
\eeq
\noindent
where 
\beq
\label{aGF}
g_i(\omega) \equiv \frac{1}{\omega - z_i \, a}
\eeq
\noindent
is referred to as {\it atomic Green's function} because it is the limit of the Green's function $G_{i \, j}$ for vanishing hopping parameter:
\beq
\label{aGF2}
\lim_{t \to 0} G_{i \, j}(\omega) =  \frac{1}{\omega - z_i \, a} \delta_{i \, j} =  g_i(\omega) \delta_{i \, j} 
\eeq
\noindent
This can easily be shown by means of (\ref{GF2}), simply by observing that $ \omega - H = \omega - a Z $ is a diagonal matrix when the hopping term vanishes.
Since the chain of adjacency matrix's elements $ A_{i \, h_1} \, A_{h_1 \, h_2} ... \, A_{h_{k-1} \, j}$ in (\ref{GF4}) is non-vanishing only if  $\{i, h_1, h_2,...,h_{k-1},j\}$ is a connected $k$-step path joining the sites $i$ and $j$, the Green's function $G_{i \, j}(\omega)$ can be viewed as a weighed sum over the paths ${\cal P}_{i\,j}$ joining $i$ and $j$:
\beq
\label{GF5}
G_{i \, j}(\omega)= g_j \sum_{{\cal P}_{i\,j}} \prod_{(h,k) \in {\cal P}_{i\,j}} p_{h\,k}
\eeq
\noindent
where $p_{h\,k}(t,\,a)= t\,g_h(a)$ is the contribution relevant to the step $(h\,k)$. Due to the extreme simplicity of the present model this quantity  is a function only of the site from which the step is taken. 
In a more refined model the hopping parameter would be a link variable, $t_{h\,k}$, and this explains the dependence of $p_{h\,k}$ on both of the sites involved in the step. Note that $p_{h\,k}$ is an adimensional quantity.
 Thus the LDOS is a local quantity, in the sense that it refers to a single site of the structure, but it takes into account the whole structure. This because it depends on a Green's function, which is a weighed sum over all possible closed paths passing through the site under examination. 
The function 
\beq
\label{Gamma}
\Gamma_{i \, j}(\omega)= \sum_{{\cal P}_{i\,j}} \prod_{(h,k) \in {\cal P}_{i\,j}} p_{h\,k}  
\eeq
\noindent
appearing in (\ref{GF5}) is related to the generating function for a random walk (RW) with probability $- a \, g_h$ for the step $(h \, k)$: $\Gamma_{i \, j}(\omega)= \tilde P_{i \, j}(-t/a)$ \cite{rwB1,rwB2}.
\noindent
The function
\beq
\label{gamma}
\gamma_{i \, j}(\omega)= \sum_{{\cal F}_{i\,j}} \prod_{(h,k) \in {\cal F}_{i\,j}} p_{h\,k},
\eeq
\noindent
built using only paths constrained as described in Section \ref{graphsS}, is in turn related to the generating function for the probability of first arrival: $\gamma_{i \, j}(\omega)=\tilde F_{i\,j}(-t/a)$. The simple RW is recovered by setting $\omega = 0$. The relation between $\Gamma_{i\,j}$ and $\gamma_{i\,j}$ is the same as the one between the generating functions of RW $\tilde P_{i\,j}$ and $\tilde F_{i\,j}$
\beq
\label{Ggamma}
\Gamma_{i \, j}(\omega)= \delta_{i \, j} + \gamma_{i \, j}(\omega) \Gamma_{j \, j}(\omega)
\eeq
\noindent
When the last site is the same as the first this reads
\beq
\label{Ggamma2}
\Gamma_i(\omega)= \frac{1}{1-\gamma_i (\omega)}
\eeq
\noindent
where $\Gamma_i(\omega)\equiv \Gamma_{i \, i}(\omega)$ and  $\gamma_i(\omega)\equiv \gamma_{i \, i}(\omega)$.

\section{Bud-Reduction}
\label{brS}
The analogous of formula (\ref{Ggamma2}) in the framework of RW is well known and is due to the factorization properties of the generating functions $\tilde P_{i\,i}$ and $\tilde F_{i\,i}$. Since the proof line for the bud-reduction technique is very similar, in the following (\ref{Ggamma2}) is proved  employing a path technique. 
 First of all we label all the relevant paths in some definite though arbitrary sequence for enumerating purposes. Thus let ${\bf P}_i=\{{\cal P}^{(n)}_{i\,i}\}_{n=1}^N$ be the set of all unrestricted paths starting and ending at site $i$, and let ${\bf F}_i=\{{\cal F}^{(n)}_{i\,i}\}_{n=1}^{\tilde N}$ be the subset of ${\bf P}$ consisting of the paths restricted as explained in Section \ref{graphsS}.
 Note that both $N$ and $\tilde N$ may be possibly infinite.
Let $\bowtie$ denote the operation of joining two paths. The only requirement for this operation is that the starting site of the second path is the same as the ending point of the first one. For instance   ${\cal P}_{i\,j} \bowtie {\cal P}_{j\,k}$ is a path starting at $i$ and ending at $k$. Let  
\beq
\label{pcont}
p({\cal P}) = \prod_{(h\,k)\in {\cal P}} t\, g_h
\eeq
be a function of paths. The following equality holds:
\beq
\label{jpcont}
p({\cal P}_{i\,j} \bowtie {\cal P}_{j\,k} ) = p({\cal P}_{i\,j}) p({\cal P}_{j\,k} ) 
\eeq
\noindent
 As already pointed out ${\bf F}_i \subset {\bf P}_i$. Moreover any concatenation of elements in ${\bf F}_i$ is in ${\bf P}_i$. More precisely for any ${\cal P}^{(h)}_{i\,i} \in {\bf P}_i$ there exists a subset of ${\bf F}_i$, $\{{\cal F}^{(k)}_{i\,i}\}_{k=1}^n$, such that ${\cal P}^{(h)}_{i\,i} = {\cal F}^{(1)}_{i\,i} \bowtie {\cal F}^{(2)}_{i\,i} \bowtie ... \bowtie {\cal F}^{(n)}_{i\,i}$. Here $n$ is nothing but the number of times the path ${\cal P}^{(h)}_{i\,i}$ reaches site $i$. Note that the elements of the subset of ${\bf F}_i$ are not necessarily all distinct. Thus, for any $h$, $p({\cal P}^{(h)}_{i\,i})$ is the product of  of factors of the  $p({\cal F}^{(k)}_{i\,i})$ kind, each raised to some positive integer power $n(k)$. At this point it is easy to understand that the following formal equalities, which yield (\ref{Ggamma2}), hold:
\beq
\Gamma_i = \sum_{k=1}^N p({\cal P}^{(k)}_{i\,i})= \sum_{h=0}^{\infty} \left[\sum_{l=1}^{\tilde N} p({\cal F}^{(l)}_{i\,i})\right]^h =\sum_{h=0}^{\infty}  \gamma_i^h = \frac{1}{1-\gamma_i}
\nonumber
\eeq

Now let us focus on the branching point. Let $*$ be a branching point splitting the graph ${\cal G}$ into two branches as in (\ref{branches}). 
In the following we are going to show that the evaluation of the Green's function at any site of one of the branches can be equivalently performed by pruning out the other branch, provided that the atomic GF at the branching point is changed into an appropriate function. It will turn out that the appropriate new atomic GF is nothing but the GF of the pruned branch at the branching site. In formula:
\beq
\label{budred}
G_{i\,i}=\left.G_{i\,i}^{\cal T}\right|_{g_* = G_{*\,*}^{\cal B}}
\eeq
where ${\cal T}$ and ${\cal B}$ denote respectively the branch containing $i$, which we call the {\it trunk} of the structure, and the branch off-springing from the branching site $*$. As superscripts, these letters denote restriction to the relevant substructure.
First of all we consider the GF at the branching site $*$. Since the paths involved in $\gamma_{*\,*}$ never go through $*$ at an intermediate step, they take place entirely either on ${\cal T}$ or on ${\cal B}$. Thus  $\gamma_{*\,*}$ is the sum of two terms:
\beq
\label{gamma*}
\gamma_*=\sum_{{\cal P}_{*\,*} \in {\cal T}} p({\cal P}_{*\,*}) + \sum_{{\cal P}_{*\,*} \in {\cal B}} p({\cal P}_{*\,*}) = \gamma_*^{\cal T} +\gamma_*^{\cal B}
\eeq
From each $p({\cal P}_{*\,*})$ appearing in (\ref{gamma*}) it is possible to factor out a term $g_*$. Thus $g_*$ appears in $\gamma_{*\,*}$,  $\gamma_{*\,*}^T$ and  $\gamma_{*\,*}^B$ only as an overall factor. This means that $g_*$ can be changed into another function, say $\bar g_*$, through the simple multiplication by the factor $g_*^{-1}\,\bar g_*$. For instance 
\beq
\label{changeg}
\gamma_*(\omega)|_{g_* = \bar g_*}= g_*^{-1}\,\bar g_* \gamma_*(\omega)
\eeq
and similarly for any $\gamma$-like function. Such a substitution of the atomic GF can also account for the introduction of an on-site potential $v_i$. In this case the Hamiltonian (\ref{GTBH}) would have a further diagonal term $V_{i\,j}=\delta_{i\,j}$ and formula (\ref{aGF}) would read 
\begin{displaymath}
\bar g_i(\omega) = \frac{1}{\omega - z_i \, a - v_i}\nonumber
\end{displaymath}
Now, making use of (\ref{GF5}) and (\ref{Ggamma2}) it is possible to write
\beq
\label{Gamma*}
G_{*\,*}= \frac{g_*}{1- \gamma_*^{\cal B} - \gamma_*^{\cal T} }=\frac{g_*}{ 1- \gamma_*^{\cal B}} \left[1-g_*^{-1}  \left(\frac{g_*}{ 1- \gamma_*^{\cal B}}\right) \gamma_*^{\cal T} \right]^{-1}
\eeq
When we restrict the structure to ${\cal T}$ or ${\cal B}$ the Green's functions at $*$ read respectively
\beq
\label{RGamma*}
G_{*\,*}^{\cal T}= \frac{g_*}{1- \gamma_*^{\cal T} } \qquad G_{*\,*}^{\cal T}= \frac{g_*}{1- \gamma_*^{\cal T}}
\eeq
Thus, recalling (\ref{changeg}),  (\ref{Gamma*}) becomes
\beq
\label{Gamma*2}
G_{*\,*}= G_{*\,*}^{\cal B}\left[1-g_*^{-1}  G_{*\,*}^{\cal B}  \gamma_*^{\cal T} \right]^{-1} =G_{*\,*}^{\cal B}\left[1-\left. \gamma_*^{\cal T}\right|_{g_* = G_{*\,*}^{\cal B}} \right]^{-1}= \left.G_{*\,*}^{\cal T}\right|_{g_* = G_{*\,*}^{\cal B}}
\eeq
which  is nothing  but (\ref{budred})  in the  special case  $i=*$. The proof for the general case can be obtained in the same fashion, and it is illustrated in some detail in Section \ref{appx}.
Formula (\ref{budred})  holds for any  branching site of  the trunk.
Thus the GFs at the sites of the trunk can be evaluated by pruning off
all the branches, provided that  the atomic GFs at the branching sites
are changed into  the corresponding GFs of the  pruned branches. Since
they carry all the  information relevant to the corresponding branches
these new atomic  GFs can be viewed as their  buds.  This operation of
bud-reduction can  be repeated  iteratively within  the  branches and
gives rise to  a   topological   simplification  of   the   original
structure. Of  course for an exact  result the knowledge  of the exact
Green's functions of the pruned  branches is still needed. When these
functions are not known a trial expression can  be plugged in, in order to 
get approximate results.
From a {\it down-top} point of view
this  approach seems
very  well suited  to engineering: the  effect, in  the
respect of the LDOS, of appending a secondary structure to a known one
is  simply obtained  by plugging  in the  appropriate atomic  GF. 
Since we will examine only the GFs relevant to the same site, from now on  we simplify the notation by dropping one of the repeated subscripts. A superscript will denote the structure relevant to the GF under examination.
The following subsections  contain a  brief survey of  possible structures
and some examples of bud-reduction.
\subsection{Fundamental Building Blocks}
In this  Subsection we apply the technique above illustrated on some simple structures, which can be used as building blocks of more complex networks.\\
$\begin{array}{cl}
\includegraphics[angle=0,width=4cm]{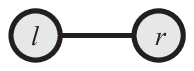}& 
\parbox{14.1cm}{\noindent {\bf Dimer}: a dimer is a simple structure whose graph consists of two sites connected by a link.
Let $g_l$ and $g_r$ be the atomic GFs at its two sites. Since there is only one restricted path starting and ending at either of its sites, it is easy to show that $\gamma_l=\gamma_r=t^2 g_l \, g_r$. Thus, for a dimer}\\
\end{array}$
\vskip 0.5cm
\beq
\label{dimer}
G_l^{\rm d}(g_l,g_r)=\frac{g_l}{1-t^2\,g_l \,g_r}
\eeq
Obviously the GF relevant to the other site has the same functional form. It is enough to swap $g_l$ and $g_r$.\\
$\begin{array}{cl}
\includegraphics[angle=0,width=4cm]{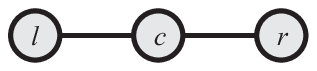}& 
\parbox{14.1cm}{\noindent {\bf Trimer}: a trimer is a simple chain whose graph consists of three sites and two links connecting two external sites to a central one. Let $g_l$ and $g_c$ and $g_r$ be the atomic GFs at its three sites. Since a trimer can be seen as a dimer branch off-springing from a dimer trunk we can write:}\\
\end{array}$
\vskip 0.5cm
\begin{eqnarray}
\label{trimer}
G_l^{\rm t}(g_l,g_c,g_r) & = & G_l^{\rm d}(g_l,G_l^{\rm d}(g_c,g_r)) =
\frac{g_l (1-t^2 g_c g_r)}{1- t^2 g_c (g_l + g_c)}\nonumber \\ 
G_c^{\rm t}(g_l,g_c,g_r) & = & G_r^{\rm d}(g_l,G_l^{\rm d}(g_c,g_r)) 
= \frac{g_c}{1- t^2 g_c (g_l + g_c)}\\
G_r^{\rm t}(g_l,g_c,g_r) & = & G_r^{\rm d}(G_r^{\rm d}(g_l,g_c),g_r)=
\frac{g_r (1-t^2 g_l g_c)}{1- t^2 g_c (g_l + g_c)}\nonumber
\end{eqnarray}
$\begin{array}{cl}
\includegraphics[angle=0,width=4cm]{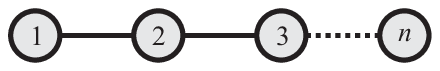} & 
\parbox{14.1cm}{\noindent {\bf $n$-mer}: iterating the procedure used for the trimer it is possible to deal with a chain of length $n$, for any $n$. More generally the atomic GF at one end of a $n$-mer can be obtained from the same function relevant to an $(k)$-mer, $k<n$:}\\
\end{array}$
\vskip 0.5cm
\beq
G_1^{(n)}(g_1,...,g_n)=G_1^{(k)}(g_1,...,g_{k-1},G_1^{(n-k+1)}(g_k,g_{k+1},...,g_n))
\eeq
In particular, when $k=2$
\beq
G_1^{(n)}(g_1,...,g_n)=G_l^{\rm{d}}(g_1,G_{0\,0}^{(n-1)}(g_2,...,g_{n-1}))=\frac{g_1}{1-t^2\,g_1\,G_1^{(n-1)}(g_2,...,g_{n-1})}
\eeq
The GF at an intermediate site can be obtained combining the results for dimers and $n$-mers as well:
\beq
G_k^{(n)}(g_1,...,g_n)=G_l^{{\rm d}}(G_k^{(k)}(g_1,...,g_k),G_1^{(n-k)}(g_{k+1},...,g_n)) \qquad \forall \, 0<k<n
\eeq
$\begin{array}{cl}
\includegraphics[angle=0,width=4cm]{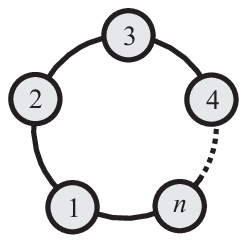}& 
\parbox{14.1cm}{\noindent {\bf Ring}: an $n$-site ring  can be obtained from $n$-mer by linking the first and the last sites. It is equivalent to an infinite linear chain obtained by repeating infinitely many times the same sequence $ \{g_1,...,g_n\}$ of the ring's atomic GFs. The periodicity constraint can be satisfied as follows: the site under examination is identified with the central site of a $(2n+1)$-mer, whose external atomic GFs are set to be equal to the one we are interested in:}
\end{array}$
\vskip 0.5cm
\beq
G_1^{{\rm R} n}(g_1,...,g_n) \equiv \bar g = G_{n+1}^{(2n+1)}(\bar g,g_2,g_3,...,g_n,g_1,g_2,g_3,...,g_n,\bar g)
\eeq
Note that the periodicity constraint introduces recursion. As it will become clearer below, this recursion forces the GF to satisfy a second order algebraic equation. This kind of equation has in general two solutions, of which only one is acceptable.  The choice criterion is given by relation (\ref{aGF2}): the GF at every site must coincide with the atomic GF at the same site for vanishing hopping parameter.

Also note that a linear homogeneous chain can be obtained as a ring of length $n$ ($n \geq 3$) whose atomic GFs are all equal.
\vskip 0.5cm
\noindent
$\begin{array}{cl}
\includegraphics[angle=0,width=4cm]{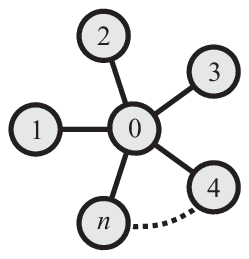}& 
\parbox{14.1cm}{\noindent {\bf junction}: the central site of a trimer can be seen as a 2-fold junction joining the two external sites. A 3-fold junction can be simply obtained by plugging a dimer GF into the atomic GF of the 2-fold junction site:}\\
\end{array}$
\vskip 0.5cm
\beq
G_0^{3\rm{j}}(g_0,g_1,g_2,g_3)=G_c^{\rm{t}}(g_1,G_{l\,l}^{\rm{d}}(g_0,g_3),g_2)=\frac{g_0}{1-t^2\,g_0\,(g_1+g_2+g_3)}
\eeq
The GF at one of the three external sites can be obtained from the one at one end of a dimer by plugging the GF of the central site of a trimer into the atomic GF of the other end:
\beq
G_1^{3\rm{j}}(g_0,g_1,g_2,g_3)=G_l^{\rm{d}}(g_1,G_c^{\rm{t}}(g_2,g_0,g_3))=\left[ 1-t^2\,g_0\,(g_2+g_3)\right]\frac{g_1}{1-t^2\,g_0\,(g_1+g_2+g_3)}
\eeq
An $n$-fold junction, for any $n$, can be obtained in exactly the same way, simply by iterating the procedure shown above. 
\begin{eqnarray}
\label{nj}
G_0^{n\rm{j}}(g_0,g_1,...,g_n) & = & G_0^{(n-1)\rm{j}}(G_l^{\rm{d}}(g_0,g_n),g_1,...,g_{n-1})=\frac{g_0}{1-t^2\,g_0\,\sum_{j=1}^n g_n} \nonumber \\\\
G_1^{n\rm{j}}(g_0,g_1,...,g_n) & = & G_l^{\rm{d}}(g_1, G^{(n-1)\rm{j}}(g_0,g_2,g_3,...,g_n)= g_1(1-t^2\, g_1 \, G_0^{n\rm{j}}) \nonumber
\end{eqnarray}
\begin{figure}[h]
\begin{center}
\includegraphics[angle=0, width=15cm]{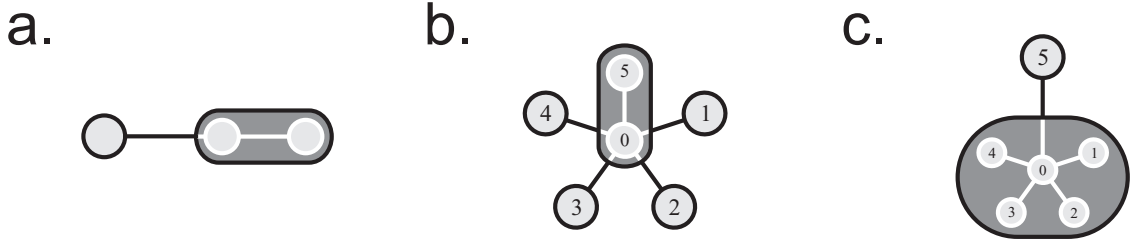} 
\caption{Examples of simple bud-reduction. The dark grey bubbles represent the bud atomic GFs. The white contoured structures inside them are the relevant branches. Note that the new atomic GF are always evaluated at the branching site; {\bf a.} Trimer. The two sites on the right site are reduced to a bud. The trimer turns into a dimer. The GF at the left and right site of this dimer give respectively the GF at the left and at the central site of the original trimer;  {\bf b.} 5-junction. The dimer consisting of the central site and site 5 is collapsed to its bud. This way it is possible to evaluate the GFs at the origin or at one of the other four peripheral sites by means of the relevant GFs for a 4-junction; {\bf c.} 5-junction. The 4 junction consisting of the origin and of the peripheral sites from 1 to 4 is collapsed to its bud. The GF for the dimer allows to evaluate the GF at site 5.}
\label{mon_junF}
\end{center}
\end{figure}
\pagebreak
\subsection{Infinite Chains}
\noindent
$\begin{array}{cl}
\includegraphics[angle=0,width=4cm]{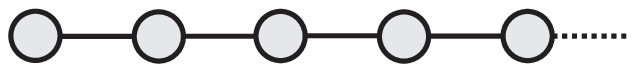}& 
\parbox{14.1cm}{\noindent {\bf Semi-infinite homogeneous chain}	 (SHC): it is a semi-infinite linear chain with constant atomic GF, $g$. The GF at its origin can be evaluated by regarding the structure as a SHC branching off from dimer trunk:}\\
\end{array}$
\vskip 0.5cm
\beq
G_0^{\rm SHC}(g)=G_l^{\rm d}(g,G_0^{\rm SHC}(g)) 
\quad \Rightarrow \quad t^2 \, g \, \left[G_0^{\rm SHC}\right]^2 - G_0^{\rm SHC}  +g =0 
\eeq
As pointed out while discussing rings, recursion due to periodicity gives rise to a second order equation. The requirement $\lim_{t \to 0} G_{0\,0}^{\rm SHC}=g$ selects the solution
\beq
G_0^{\rm SHC}(g)=\frac{1-\sqrt{1-(2 \,t\,g)^2}}{2\,t^2 \,g} =\frac{2 g}{1+\sqrt{1-(2 \,t\,g)^2}}
\eeq
The GF at a site different from the origin is evaluated as the same function for one of the ends of an $n$-mer, provided that the atomic GF at that end is changed into the GF at the origin of a SHC.  \vskip 0.5cm
\noindent
$\begin{array}{cl}
\includegraphics[angle=0,width=4cm]{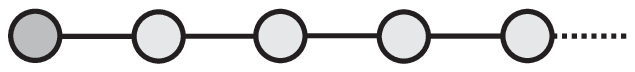}& 
\parbox{14.1cm}{\noindent {\bf Semi-infinite chain} (SC): in a semi-infinite chain the origin is a special site; since it is the only site with coordination 1, it may have a different atomic GF. Let therefore be $g_0$ the atomic GF at the origin and $g$ the atomic GF at any other site. }\\
\end{array}$
\vskip 0.5cm
\beq
\label{SC}
G_0^{\rm SC}(g_0,g)=G_l^{\rm d}(g_0,G_0^{\rm SHC}(g))= \left[ \frac{1}{g_0}-\frac{1}{2\,g} \left( 1-\sqrt{1-(2 \,t\,g)^2}\right)\right]^{-1}
\eeq
The procedure for a site different from the origin is the same as the one outlined for the SHC. Both in SC and SHC cases (\ref{SC}) provides an alternative way: the GF at the $k$-th site is the same as the one for the origin of a SC provided that the atomic GF at the origin is the GF for the end of an appropriate $k$-mer.\vskip 0.5cm
\noindent
$\begin{array}{cl}
\includegraphics[angle=0,width=4cm]{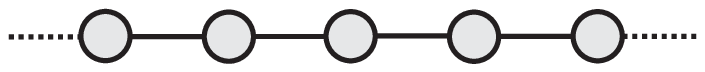}& 
\parbox{14.1cm}{ \noindent {\bf Homogeneous infinite chain} (HIC): let $g$ be the atomic GF at each of the equivalent sites of an infinite chain. We can obtain the GF at one of these sites by regarding it as a branching site where a SHC branch springs off from a SC trunk. Alternatively the whole structure may be regarded as a dimer trunk carrying two SHC branches:}\\
\end{array}$ \vskip 0.5cm
\beq
\label{HIC}
G_0^{\rm HIC}(g)=G_0^{\rm SC}\left(G_0^{\rm SHC}(g),g\right)=G_l^{\rm d}(G_0^{\rm SHC}(g),G_0^{\rm SHC}(g))= \frac{g}{\sqrt{1-(2 \,t\,g)^2}}
\eeq
This structure is equivalent to a ``ring'' composed of two equivalent sites. Thus the same result as in (\ref{HIC}) can be obtained from a recursive second order equation for a trimer. The required GF is the GF at the central site of a trimer in whose external atomic GFs we plug the required result itself. The central atomic GF is the same as in the original structure. \vskip 0.5cm
\noindent
$\begin{array}{cl}
\includegraphics[angle=0,width=4cm]{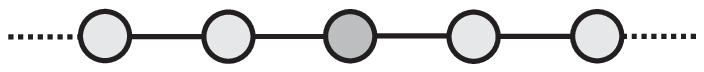}& 
\parbox{14.1cm}{\noindent {\bf Infinite chain with impurity} (ICI): if we change the atomic GF of just one site of the HIC above into $g_0$ we obtain an infinite chain with a single impurity. The GF at the impurity site can be obtained as follows:}\\
\end{array}$
\vskip 0.5cm
\begin{eqnarray}
\label{ICI}
G_0^{\rm ICI}(g_0,g) & = & G_0^{\rm SC}\left(G_0^{\rm SC}(g_0,g),g\right)=G_l^{\rm d}(G_0^{\rm SC}(g_0,g),G_0^{\rm SHC}(g)) \nonumber \\
& = & G_c^{\rm t}(G_0^{\rm SHC}(g),g_0,G_0^{\rm SHC}(g)) = \left[ \frac{1}{g_0} - \frac{1}{g} \left(1-\sqrt{1-(2 \,t\,g)^2}\right) \right]^{-1}
\end{eqnarray}
\subsection{Bethe Lattices and Cayley Trees}
\label{BLCTS}
\begin{figure}[h]
\begin{center}
\includegraphics[angle=0, width=15cm]{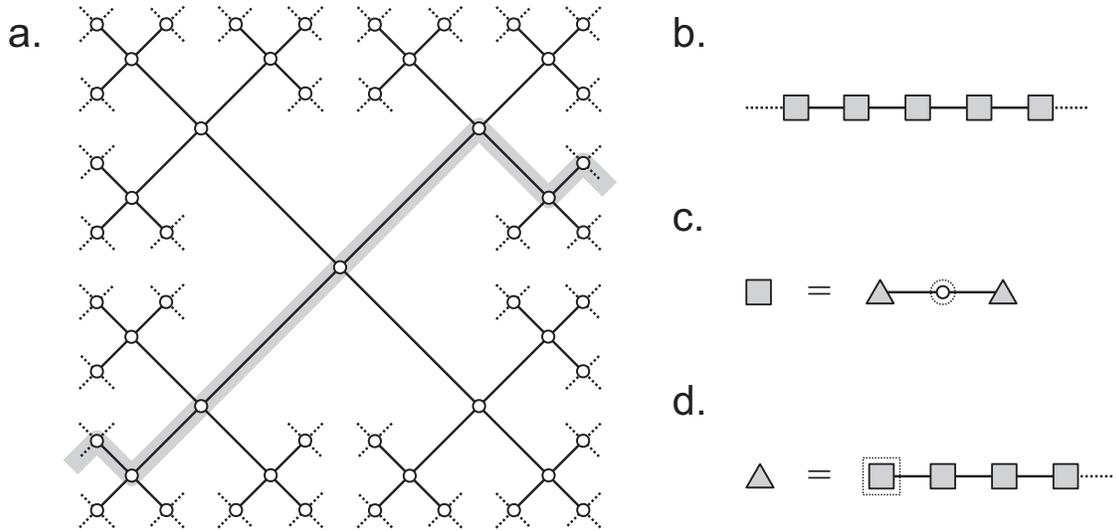} 
\caption{Four fold Bethe lattice (4-Bl); {\bf a.} The whole structure is shown. The links are drawn with different lenghts just for graphical convenience. The shading shows one possible chain substructure which can be chosen as the trunk for the bud-reduction; {\bf b.} Bud-reduced structure: only the trunk is shown.  {\bf c.} Bud-reduced structure: the structure of the main buds is shown. They consist of trimers whose external sites' GF are the result of a further bud-reduction.  {\bf d.} Bud-reduced structure: the structure of the secondary buds is shown. Note that, due to the recursivity of the structure, they are defined in terms of the main buds. The sites at which the GFs are evaluated are surrounded by dashed lines. }
\label{nBlF}
\end{center}
\end{figure}
\begin{figure}[h]
\begin{center}
\includegraphics[angle=0, width=15cm]{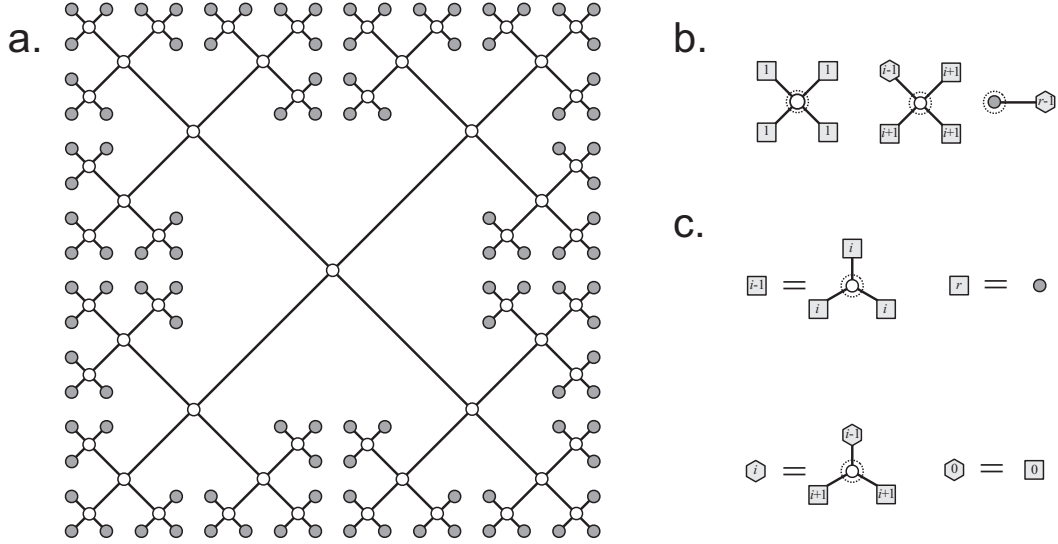} 
\caption{Four fold Cayley tree ((4,4)-Ct); {\bf a.} The whole structure is shown. The links are drawn with different lenghts for graphical convenience. The sites with coordination $z=1$ are shaded in grey. 
 {\bf b.} Bud-reduced structures for the GF of the structure. From left to right: site at the origin, site at distance $i$ from the origin, peripheral site (at distance $r$ from the origin); {\bf c.} recursion relations for the bud-reduced structures. 
The sites at which the GFs are evaluated are surrounded by dashed lines.}
\label{nCtF}
\end{center}
\end{figure}

 
An $n$-fold Bethe lattice ($n$-Bl) is an infinite network of sites with coordination $n$, i.e. of $n$-fold junctions. Note that the HIC is nothing but a 2-Bl. Actually a Bl is the simplest looples homogeneous structure after the homogeneous chain. As in a homogeneous chain all the sites of the $n$-Bl are equivalent.
A possible way to work out the GF at one of the $n$-Bl's sites consists in considering any HIC subgraph of this structure as its trunk. Inside each of the $n$-2 branches off-springing at every site of this trunk one can recognize in turn a SHC trunk, dressed with the same atomic GF as the main trunk. 
Thus the $n$-Bl can be seen as a HIC, whose atomic GF, $J$, is the one for the central site of a  ($n$-2)-junction; the atomic GF, $P$  for the peripheral sites of this junction are in turn the GF for the origin of a SHC, whose atomic GFs once again equal $J$. Equations (\ref{nBl}) and Figure~\ref{nBlF}.
display the bud-reduction described above:
\beq
\label{nBl}
G_i^{n{\rm Bl}}= G_0^{\rm HIC}\left(J(g)\right) \qquad J(g)=G_0^{\rm (n-2){\rm j}}(g,\underbrace{S(g),S(g),...}_{n-2\;{\rm entries}}) \qquad S(g)= G_0^{\rm SHC}\left(J(g)\right)
\eeq
The combination of the last two equations of (\ref{nBl}), together with (\ref{nj}) and (\ref{SC}), give a self-consistency equation for $J$. The GF for the $n$-Bl is thus worked out by plugging the  expression obtained for $J$ into  (\ref{HIC}), as required by the first of (\ref{nBl}): 
\beq
\label{GnBl}
G_i^{n{\rm Bl}}(g)=\frac{2 (n-1) g}{n - 2 + n \sqrt{1-(n-1)(2\, g\, t)^2}}
\eeq
The details of this calculation are given in Section \ref{appx}.
As stated above a Bethe Lattice is an infinite homogeneous structure, and its sites are all equivalent. Its finite version is known as {\it Cayley tree} \cite{dend2}. A $n$-fold Cayley tree of radius $r$ ($(n,r)$-Ct) is a structure invariant under $n$-fold discrete rotations centered at its central site, which we refer to as its origin.  Every site whose distance from the origin is lower than the radius has coordination number $z = n$. The peripheral sites, whose distance from the origin equals the radius, have coordination number $z = 1$. A $(n,r)$-Ct is thus a non-homogeneous structure. Only the sites placed at the same distance from the origin are equivalent. Note that, due to the exponential growth, the number of peripheral sites is of the same order as the number of internal sites. As illustrated by Fig.~\ref{nCtF}
the GF at the origin of a $(n,r)$-Ct can be obtained by means of three bud-reduction equations:
\beq
\label{Ctbr}
G_0^{(n,r){\rm Ct}}(g_n) = G_0^{\rm n{\rm j}}(g,\underbrace{L_1(g_n),L_1(g_n),...}_{n \;{\rm entries}}) \qquad L_i(g_n)= G_0^{\rm (n-1){\rm j}}(g_n,\underbrace{L_{i+1}(g_n),L_{i+1}(g_n),...}_{n-1 \;{\rm entries}}) \qquad L_r = g_1
\eeq
where $g_n$ is the atomic GF at any of the internal sites ($z=n$), while $g_1$ is the  atomic GF at any of the peripheral sites ($z=1$). Note that the second of the (\ref{Ctbr}) sets a recursion relation which is closed by the third equation. Also note that imposing the fixed point condition $L_i(g_n) = L_{i+1}(g_n)=L(g_n)$ and plugging the solution into the first equation of  (\ref{Ctbr}) one recovers the result for the $n$-fold Bethe lattice, formula (\ref{GnBl}). As illustrated in Figure \ref{nCtF}
, the bud-reduction for the sites other than the origin is a little more complex, due to the lower degree of symmetry. 
\subsection{Fern Lattices}
A fern is a plant characterized by an evident self-similarity. 
The secondary structures branching off from the the main stalk mimic the whole structure. This resemblance between the branches and the main structure goes on up to a certain order $f$, at which the branches are simple leaves or stalks.
\begin{figure}[h]
\begin{center}
\includegraphics[angle=0, width=15cm]{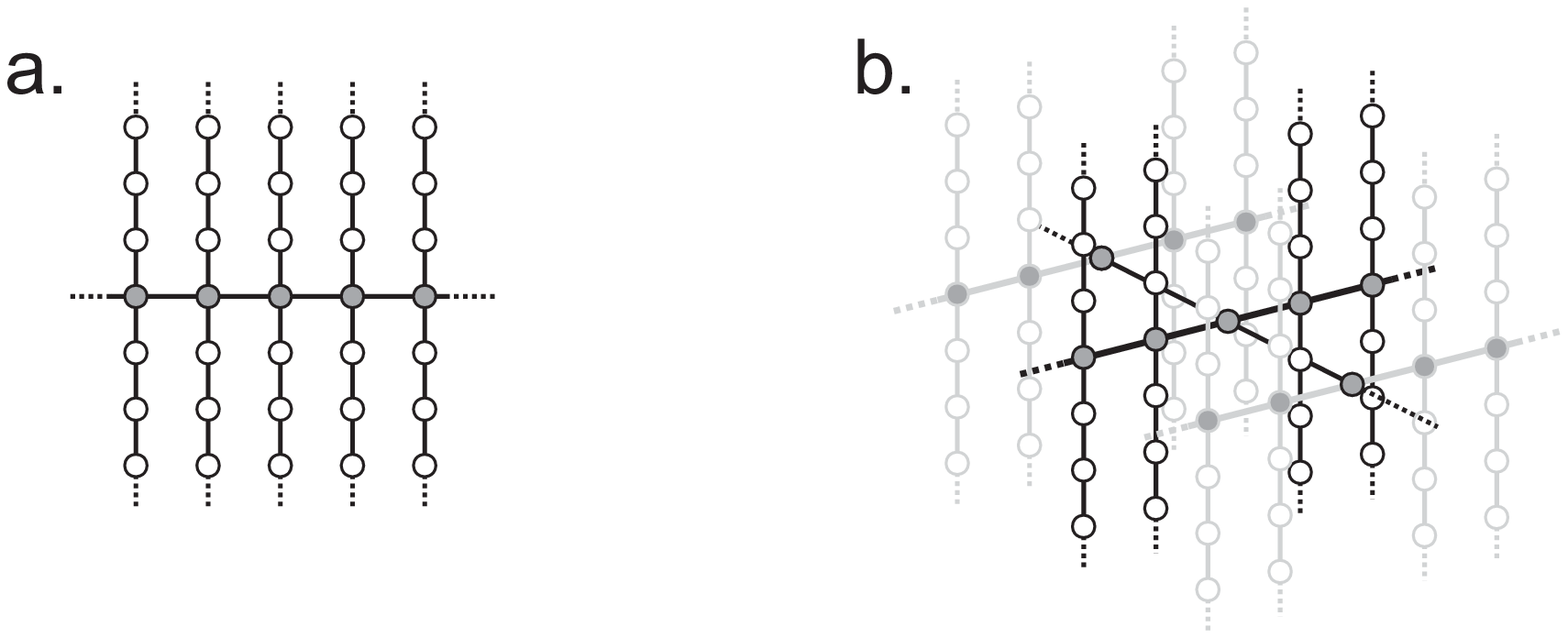} 
\caption{Fern lattices. Filled circles: sites with $z=4$. Unfilled circles: sites with $z=2$; {\bf a.} $(2,1)$ fern lattice, also known as comb lattice or fishbone; {\bf b.} $(2,2)$ fern lattice; some secondary structures are drawn in light grey only for graphical convenience. }
\label{fernF}
\end{center}
\end{figure}
An infinite {\it fern lattice} has a similar structure: at every site of an infinite homogeneous chain $b$ identical secondary structures branch off. These are in turn semi-infinite homogeneous chains carrying $b$ identical tertiary structures, and so on. At some order $f$ the $b$ substructures branching off from every site are simple semi-infinite homogeneous chains (see Figure \ref{fernF}). Thus a fern lattice is characterized by two parameters, the {\it branching number}, $b$, and the {\it order}, $f$. Most of the sites of a fern lattice, namely the ones on the branches of highest order, have coordination number $z=2$. All the other ones have coordination number $z=b+2$.
 Due to translational invariance along the main stalk, the sites of a $(b,f)$ fern ($(b,f)$F) are naturally divided into equivalence classes distinguished by two parameters: the order $i$ of the stalk on which they are placed and the distance $d$ from the stalk of order $i-1$. Thus $i$ ranges from 0, on the main stalk, to $f$, on the ``leaves'', whereas $-\infty \leq d \leq \infty$ if $i=0$ and $d \geq 1$ if $i>0$. The GF for the sites of a $(b,f)$F can be obtained as
\beq
\label{fern1}
G_{(0,d)}^{(b,f){\rm F}}(g_s,g_l)= G_d^{\rm HIC}\left(J(g_s,L_1)\right) \qquad 
G_{(i,d)}^{(b,f){\rm F}}(g_s,g_l)= G_d^{\rm SIC}\left(O_i,J(g_s,L_i)\right) 
\eeq
where $g_s$ is the atomic GF at the sites with coordination number $z=b+2$, whereas $g_l$ is the same quantity for the sites with coordination number $z=2$. The other functions appearing in (\ref{fern1}) are defined by the relations: 
\beq
\label{fern2}
J(g_s,L_i)=G_0^{\rm b{\rm j}}(g,\underbrace{L_i,L_i,...}_{b\; {\rm entries}}) \qquad O_i=G_0^{\rm (b+1){\rm j}}(g_s,L_{i-1}, L_{i-1},\underbrace{L_i,L_i,...}_{b-1\; {\rm entries}})
\eeq
and
\beq
\label{fern3}
L_i= G_0^{\rm SHC}\left(J(g_s,L_{i+1})\right) \qquad J(g_s,L_f)=g_l 
\eeq
It is interesting to note that the fixed point relation $L_{i+1}=L_i$ sets in some sense $f=\infty$. In this case all the sites are equivalent and  have coordination number $z=b+2$. Thus a $n$Bl can be thought of as a $(n-2,\infty)$ fern, and indeed the equations (\ref{fern1}), (\ref{fern2}) and (\ref{fern3}), together with the fixed point relation, give (\ref{GnBl}) once again. 
\section{Conclusions}
\label{conclS}
In this paper we face the general 	problem of determining the spectral properties of inhomogeneous structures. Since the lack of translational invariance rules out a powerful analytical tool such as the Fourier transform, and forces a direct-space approach, the spectral properties are recovered through the evaluation of the Green's functions, which are explicitly site-dependent quantities.

An analytical technique, based on the properties of the GFs in presence of branching sites, is developed and illustrated by means of a many examples. This technique, which we call bud-reduction, allows to account for an entire substructure of the network under examination by means of a sort of on-site potential of geometrical origin. It can also be performed telescopically, meaning that the bud-reduced branches may in turn carry substructures which have already undergone bud-reduction.  
This tool yields a topological simplification of the structure under examination and allows a better understanding of the effect of each substructure on the spectral properties of the system. For this reason it is very promising from the point of view of reverse engineering: the insight gained on the effects of different topologies may be a guide in designing networks displaying the desired spectral features.
\section{Appendix: Explicit Calculations}
\label{appx}
\subsection{Bud-Reduction Formula: the General Case.}
In Section \ref{brS} we proved formula (\ref{Gamma*2}), which is nothing but bud-reduction formula (\ref{budred}) in the special situation where the site under examination, $i$, coincides with the branching site, $*$. We recall that the latter is the unique intersection of the two substructure forming the network under examination. We refer to these structures as the trunk ${\cal T}$ and the branch  ${\cal B}$. The site under examination, $i$, belongs to ${\cal T}$.

The general case, (\ref{budred}), can be easily  understood in terms of paths. The function $\gamma_{i\,i}$  involves paths which  reach $i$  only after  the last
step. They  can be divided into  two sets: the ones  which never visit
$*$, which  therefore take place entirely  on ${\cal T}$, and  the ones which
visit $*$  at least  once, and  can therefore take  some steps  on ${\cal B}$
(remember that $*$ is the only ``door'' to ${\cal B}$). We denote these paths
respectively  ${\cal  F}_{i\,i}^{\not  *}$ and  ${\cal F}_{i\,i}^*$. Any path ${\cal F}_{i\,i}^*$ passing through $*$
has the same structure: it is the composition of

\begin{itemize}
\item a path ${\cal F}_{i,*}$ which takes place entirely on ${\cal T}$ and never reaches $*$ at an intermediate step.
\item a path  ${\cal P}_{*,*}$ which takes place on both  ${\cal T}$ and ${\cal B}$ and goes through $*$ an unrestricted number of times.
\item a path  ${\cal F}_{*,i}$ which  takes place only on ${\cal T}$ and reaches $i$ only after the last step, without going through $*$ any more. 
\end{itemize}
Thus 
\begin{eqnarray}
\label{array}
\gamma_i & = &\sum_{{\cal F}_{i,i}^*} p({\cal F}_{i,i}^*)+\sum_{{\cal F}_{i,i}^{\not *}} p({\cal F}_{i,i}^{\not *})=
\sum_{{\cal F}_{i,i}^*} p({\cal F}_{i,*} \bowtie {\cal P}_{*,*} \bowtie {\cal F}_{*,i}) +  \sum_{{\cal F}_{i,i}^{\not *}} p({\cal F}_{i,i}^{\not *})= \nonumber \\ \nonumber
& = &\sum_{{\cal F}_{i,*}} p({\cal F}_{i,*})  \sum_{{\cal P}_{*,*}} p({\cal P}_{*,*})  \sum_{{\cal F}_{*,i}} p({\cal F}_{*,i}) +  \sum_{{\cal F}_{i,i}^{\not *}} p({\cal F}_{i,i}^{\not *})=\\ 
&  = &\sum_{{\cal F}_{i,*}} p({\cal F}_{i,*})  G_{*\,*}\,  g_*^{-1} \sum_{{\cal F}_{*,i}} p({\cal F}_{*,i}) +  \sum_{{\cal F}_{i,i}^{\not *}} p({\cal F}_{i,i}^{\not *}) =\\ \nonumber
& = & \sum_{{\cal F}_{i,*}} p({\cal F}_{i,*})   \sum_{{\cal F}_{*,i}} p({\cal F}_{*,i})|_{ g_*= G_{*\,*}} +  \sum_{{\cal F}_{i,i}^{\not *}} p({\cal F}_{i,i}^{\not *})
\end{eqnarray}
where we used the fact that $g_*$ appears as an overall factor in $\sum_{{\cal F}_{*,i}} p({\cal F}_{*,i})$. If we restrict ourselves to the trunk ${\cal T}$ we similarly get
\beq
\gamma_i^{\cal T}=\sum_{{\cal F}_{i,*}} p({\cal F}_{i,*})   \sum_{{\cal F}_{*,i}} p({\cal F}_{*,i})|_{ g_*= G_{*\,*}^{\cal T}} +  \sum_{{\cal F}_{i,i}^{\not *}} p({\cal F}_{i,i}^{\not *})
\eeq
where $ G_{*\,*}^{\cal T} $ is nothing but the Green's function at $*$ restricted to the trunk. This means that $\gamma_i$ of the whole structure can be obtained through the same function restricted to the trunk, by the substitution $G_{*\,*}^{\cal T} = G_{*\,*}$. According to (\ref{Gamma*2}) this is accomplished by the substitution $g_* = G_{*\,*}^{\cal B}$ in $G_{*\,*}^{\cal T}$. Thus, since $g_*$ appears only into $G_{*\,*}$,
\beq
\label{budredal}
G_{i\,i}=\frac{g_i}{1-\gamma_i}=\left.\frac{g_i}{1-\gamma_i^{\cal T}}\right|_{g_* = G_{*\,*}^{\cal B}}=\left.G_{i\,i}^{\cal T}\right|_{g_* = G_{*\,*}^{\cal B}}
\eeq
which is nothing but (\ref{budred}). 
\subsection{Nested Bud-Reduction: Bethe Lattice}
Equations (\ref{nBl}) explicitly read
\beq
\label{EnBl}
G_i^{n{\rm Bl}}= \frac{J}{\sqrt{1-(2\,t\, J)^2}}
\qquad J = \frac{g}{1-(n-2)\,t^2\,g\,S}
\qquad S = \frac{1 - \sqrt{1-(2\,t\, J)^2}}{2 t^2 \,J}
\eeq
As we noted in Section \ref{BLCTS} the last two equations of (\ref{EnBl}) yield a self-consistency equation,,
\beq
2\,(J-g) = g\,(n-2)(1 - \sqrt{1-(2\,t\, J)^2}),
\eeq
whose only acceptable solution has the form
\beq
\label{BlJ}
J(g)= \frac{n\,g- (n-2)\sqrt{g^2\left[1-(n-1)(2 g\,t)^2\right]}}{2+2\left[(n-2)\,g\,t\right]^2}
\eeq
The substitution of (\ref{BlJ}) into the first of (\ref{EnBl}) gives the 
(\ref{GnBl}) through the use of the following identities:
\[
\left(n\,(n-2)\,(g\,t)^2 - R\right)^2 = \left[1+\left((n-2)\,g\,t\right)^2\right]-(g\,t)^2\left[n-(n-2) R\right]^2 
\]
\[
\frac{n-(n-2)\,R}{n\,(n-2)(g\,t)^2 - R}=\frac{n-1}{(n-2)+n R}
\]
where $R=\sqrt{1-(n-1)(2 g\,t)^2}$.
\pagebreak
\bibliography{}

\pagebreak
\listoffigures

\end{document}